\documentclass[aps,prc,twocolumn,showpacs]{revtex4}
\usepackage{amssymb}
\usepackage{amsmath}
\usepackage{graphicx}
\usepackage{lscape}
\usepackage{booktabs}
\usepackage{epsfig}
\begin{document}

\title{Comparative study for non-statistical fluctuation of \\ net- proton,
       baryon, and charge multiplicities}
\author{Dai-Mei Zhou$^1$ \footnote{zhoudm@phy.ccnu.edu.cn/zdm@iopp.ccnu.edu.cn}, Zeng-Zeng Luo$^1$,
        Yun cheng $^1$, Ayut Limphirat$^{2,3}$, Yu-Liang Yan$^4$, Yu-Peng
        Yan$^{2,3}$, Xu Cai$^1$, and Ben-hao Sa $^{1,4}$}

\affiliation{$^1$ Key Laboratory of Quark and Lepton Physics (MOE) and
              Institute of Particle Physics, Central China Normal University,
              Wuhan 430079, China. \\
             $^2$ School of Physics, Institute of Science, Suranaree
              University of Technology, Nakhon Ratchasima 30000, Thailand.\\
             $^3$ Thailand Center of Excellence in Physics (ThEP), Commission
              on Higher Education, Bangkok 10400, Thailand.\\
             $^4$ China Institute of Atomic Energy, P. O. Box 275 (10),
              Beijing, 102413 China.}

\begin{abstract}
We calculate the real and non-statistical higher moment excitation functions
($\sqrt{s_{NN}}$ = 11.5 to 200 GeV) for the net-proton, net-baryon, and the
net-charge number event distributions in the relativistic Au+Au collisions
with the parton and hadron cascade model PACIAE. It turned out that because
of the statistical fluctuation dominance it is very hard to see signature of
the CP singularity in the real higher moment excitation functions. It is
found that the property of higher moment excitation functions are
significantly dependent on the window size, and hence the CP signatures may
show only in a definite window for a given conserved observable
non-statistical higher moments. But for a given widow size, the CP singularity
may show only in the non-statistical higher moment excitation functions of a
definite conserved observable.\\
\end{abstract}
\pacs{25.75.Dw, 24.85.+p}
\maketitle

%%%%%%%%%%%%%%%%%%%%%%%%%%%%%%%%%%%%%%%
\section {Introduction}
%%%%%%%%%%%%%%%%%%%%%%%%%%%%%%%%%%%%%%%
One fundamental aim of relativistic heavy-ion collisions is to
explore the phase transition from the hadronic matter (HM) to the
quark-gluon matter (QGM). Evidences for the strongly coupled
quark-gluon plasma (sQGP) were reported years ago
\cite{brah,phob,star,phen}. The RHIC beam energy scan (BES) was then
proposed \cite{moha}. Attempts are being made to experimentally
locate the QCD critical point (CP), where the first order phase
transition gives over to the ``crossover" \cite{aoki} in the QCD
phase diagram of temperature $T$ vs. baryon chemical potential
$\mu_B$ \cite{aoki1,chen,fodo,gava}. Because a cms energy
($\sqrt{s_{NN}}$) in the heavy-ion collisions maps out a $\mu_B$
value, a $T$ value can be extracted by fitting the particle ratio in
the hadronic resonance gas (HRG) model \cite{peter,cley} to
corresponding data in the heavy-ion collisions. The scanning over
the phase diagram can then be performed. Meanwhile, the search for
CP signatures could also be done at each cms energy via the
singularity (nonmonotonicity) shown in higher moment excitation
functions (higher moment as a function of reaction energy) of the
conserved observables. Since the birth of RHIC BES considerable
progresses have been made in this field both experimentally
\cite{star1,gupta,luo,moha1} and theoretically
\cite{rajiv,scha,pnjl,dse,yang,peter1,zhim,wang}. A upsurge was
reached last year at the 7$^{th}$ International Workshop on Critical
Point and Onset of Deconfinement, Nov. 7-11 Wuhan, China. Although
many relevant fundamental physics have been demonstrated, the
mysteries remain.

Early in 2009 one employed the UrQMD model investigating the
net-proton, net-baryon, and net-charge kurtosis in the central
Pb+Pb/Au+Au collisions from $E_{lab}$=2$A$ GeV to
$\sqrt{s_{NN}}$=200 GeV \cite{blei}. They did not find any evidence
of CP singularity in the kurtosis excitation functions. However,
they did find that the kurtosis excitation functions behave quite
different among the net-proton, net-baryon, and net-charge: The
net-proton kurtosis becomes slightly negative at low $\sqrt{s_{NN}}$
and the net-baryon kurtosis decreases to a large negative value
while the net-charge kurtosis keeps zero approximately. Similarly,
the kurtosis as a function of the rapidity window size were also
different: Both the net-proton kurtosis and net-charge kurtosis are
differently fluctuated around zero, but the net-baryon kurtosis
reaches a large negative value at the lager window. On the contrary,
the opinion in \cite{asak} is quite opposite. By assuming that the
number distribution of a conserved observable is a binomial function
and the strange baryon can be approximately considered, they
analytically derived a relation between baryon number cumulant and
the proton number cumulant. This means that the net-baryon cumulant
has a similar behavior to the net-proton.

Recently it has been reported in \cite{koch} that the effect of
finite acceptance dramatically influences the cumulants of conserved
observables, which makes the interpretation of experimental data
measured in a given acceptance very uncertain. In \cite{koch} the
concept of ``required acceptance" is introduced, where all the
particles, which are necessary for the physics in question, are
covered. The required acceptance is not the full acceptance where
all particles in the final hadronic state are covered. The authors
assume that the bias of a real acceptance in a real experiment from
the required acceptance can be represented by a binomial probability
distribution, then they derive the general formula that relate the
true cumulants which reflect the full dynamics of the system
(require acceptance) with the actually measured cumulants (real
acceptance). This relation involves additional moments which cannot
be expressed by the cumulants and should be simultaneously measured in
order to extract information about CP singularity. But how to
measure these additional moments is not yet known.

We attempt, in this paper, to further explore the relevant issues
using the parton and hadron cascade model PACIAE \cite{yan}. The
net-proton, net-baryon, and the net-charge number event distributions
and their higher moment excitation functions ($\sqrt{s_{NN}}$=11.5
to 200 GeV) in the 0-5\% most central Au+Au collision are calculated
in different transverse momentum and pseudorapidity windows by the
PACIAE model. As pointed out in \cite{zhou} that the study of higher
moment excitation function singularity is a matter of dynamical
fluctuation, but the statistical fluctuation is always dominant. In
order to pronounce the effect of dynamical fluctuation,
non-statistical higher moments (cumulants) of conserved observables
are proposed in \cite{zhou}. Therefore, we calculate simultaneously
both the real and the non-statistical higher moment excitation
functions, and comparatively analyze the dependence of these
excitation functions on the window size.

%%%%%%%%%%%%%%%%%%%%%%%%%%%%%%%%%%%%%%%
\section {Models}
%%%%%%%%%%%%%%%%%%%%%%%%%%%%%%%%%%%%%%%
Suppose that the event distribution of the number of conserved
observable $x$ is $P(x)$, then its n$^{th}$ moment about zero is
\begin{equation}
\langle x^{(n)}\rangle=\int{x^nP(x)dx},
\end{equation}
and its n$^{th}$ moment about the mean ($M\equiv\langle x^{(1)}\rangle
\equiv\langle x\rangle$) is
\begin{equation}
M^{(n)}=\langle(x-\langle x \rangle)^n\rangle=\int{(x-\langle x
\rangle)^nP(x)dx}. \label{mea}
\end{equation}
Therefore its higher moments and moment products investigated widely are:
%\begin{flushleft}
\begin{equation}
{\rm variance:} \hspace{1cm} \sigma^2=M^{(2)}, \label{sig}
\end{equation}
\begin{equation}
{\rm skewness:} \hspace{1cm} S=M^{(3)}/(M^{(2)})^{3/2}, \label{ske}
\end{equation}
\begin{equation}
{\rm kurtosis:} \hspace{1cm} \kappa=M^{(4)}/(M^{(2)})^2-3,
\label{kur}
\end{equation}
%\end{flushleft}
\begin{equation}
S\sigma=M^{(3)}/M^{(2)},
\label{ssig}
\end{equation}
and
\begin{equation}
\kappa\sigma^2=M^{(4)}/M^{(2)}-3M^{(2)}.
\label{ksig}
\end{equation}

We employ the parton and hadron cascade model PACIAE to generate the real
event distributions. The mixed events are generated by combining particles
selected randomly from different real events, while reproducing the event
multiplicity distribution of the real events \cite{afan}. If the
n$^{th}$ real (mixed) moment about the mean calculated with real
(mixed) events is denoted by $M^{(n)}_R$ ($M^{(n)}_M$), then the
corresponding n$^{th}$ non-statistical moment is
\begin{equation}
M^{(n)}_{NON}=M^{(n)}_R-M^{(n)}_M,
\label{rm}
\end{equation}
like in \cite{afan}.

PACIAE is based on PYTHIA \cite{sjos} which is devised for the high energy
hadron-hadron (hh) collisions. In the PYTHIA model, a hh collision is first
decomposed into the parton-parton collisions. The hard parton-parton
interaction is dealt with lowest leading order perturbative QCD (LO-pQCD) and
the soft parton-parton interaction is considered empirically. Because the
consideration of the initial- and final- state parton showers as well as the
parton multiple interactions, the consequence of a hh collision is a partonic
state composed of quarks (anti-quarks), diquarks (anti-diquarks) and gluons.
It is followed by the string construction and fragmentation. So a final
hadronic state is obtained for a hh collision eventually.

In the PACIAE model \cite{yan}, a nucleus-nucleus collision is first
decomposed into nucleon-nucleon (NN) collisions according to the collision
geometry and NN total cross section. Each NN collision is dealt by PYTHIA
with string fragmentation switched-off and the diquarks (anti-diquarks)
broken into quark pairs (anti-quark pairs). When all NN collision pairs are
exhausted, a partonic initial state (composed of quarks, antiquarks, and
gluons) is obtained for a nucleus-nucleus collision. This partonic initial
stage is followed by a parton evolution stage, where parton rescattering is
performed by the Monte Carlo method with $2\rightarrow2$ LO-pQCD cross
sections \cite{BL1977}. The hadronization stage follows the parton
rescattering stage. The Lund string fragmentation model and a phenomenological
coalescence model are provided for the hadronization. Then the rescattering
among the produced hadrons is dealt with the usual two body collision model
\cite{yan}. In this hadronic evolution stage, only the rescatterings among
$\pi$, $K$, $p$, $n$, $\rho (\omega)$, $\Delta$, $\Lambda$, $\Sigma$, $\Xi$,
$\Omega$, and their antiparticles are considered for simplicity. We refer to
\cite{yan} for the details.
\begin{figure}
\centering
\includegraphics[width=3.in]{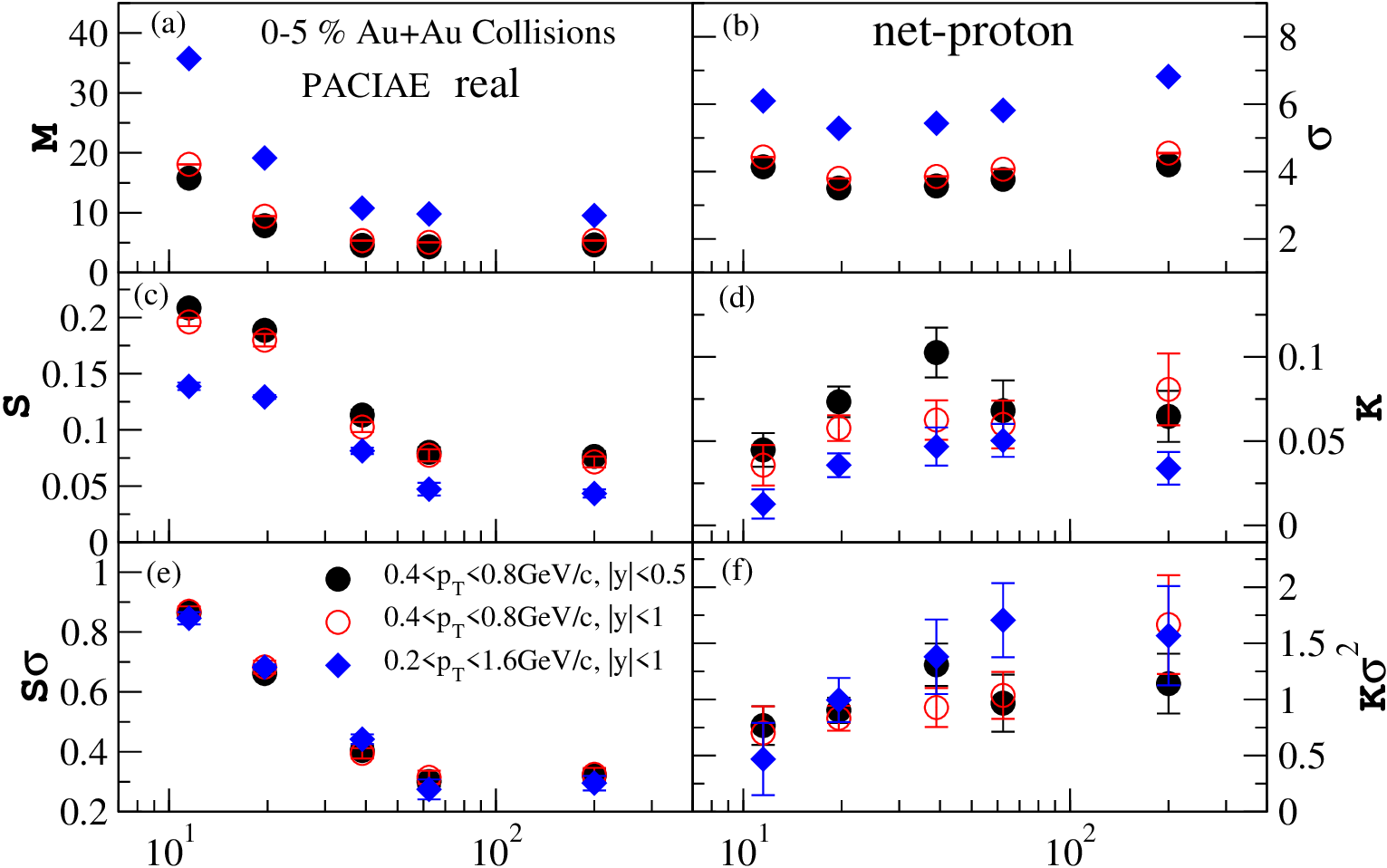}
\vskip 1.0cm
\includegraphics[width=3.in]{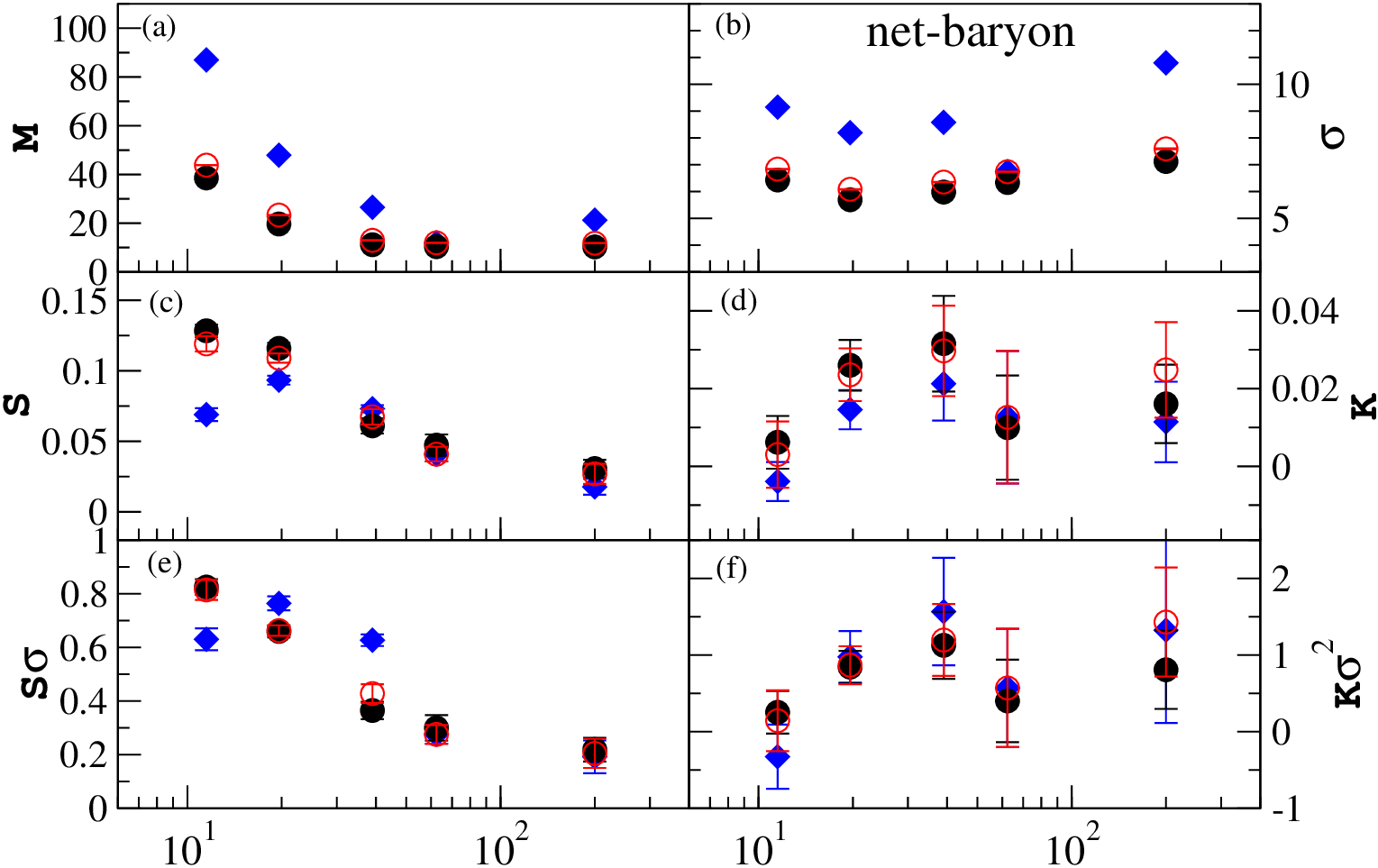}
\vskip 1.0cm
\includegraphics[width=3.in]{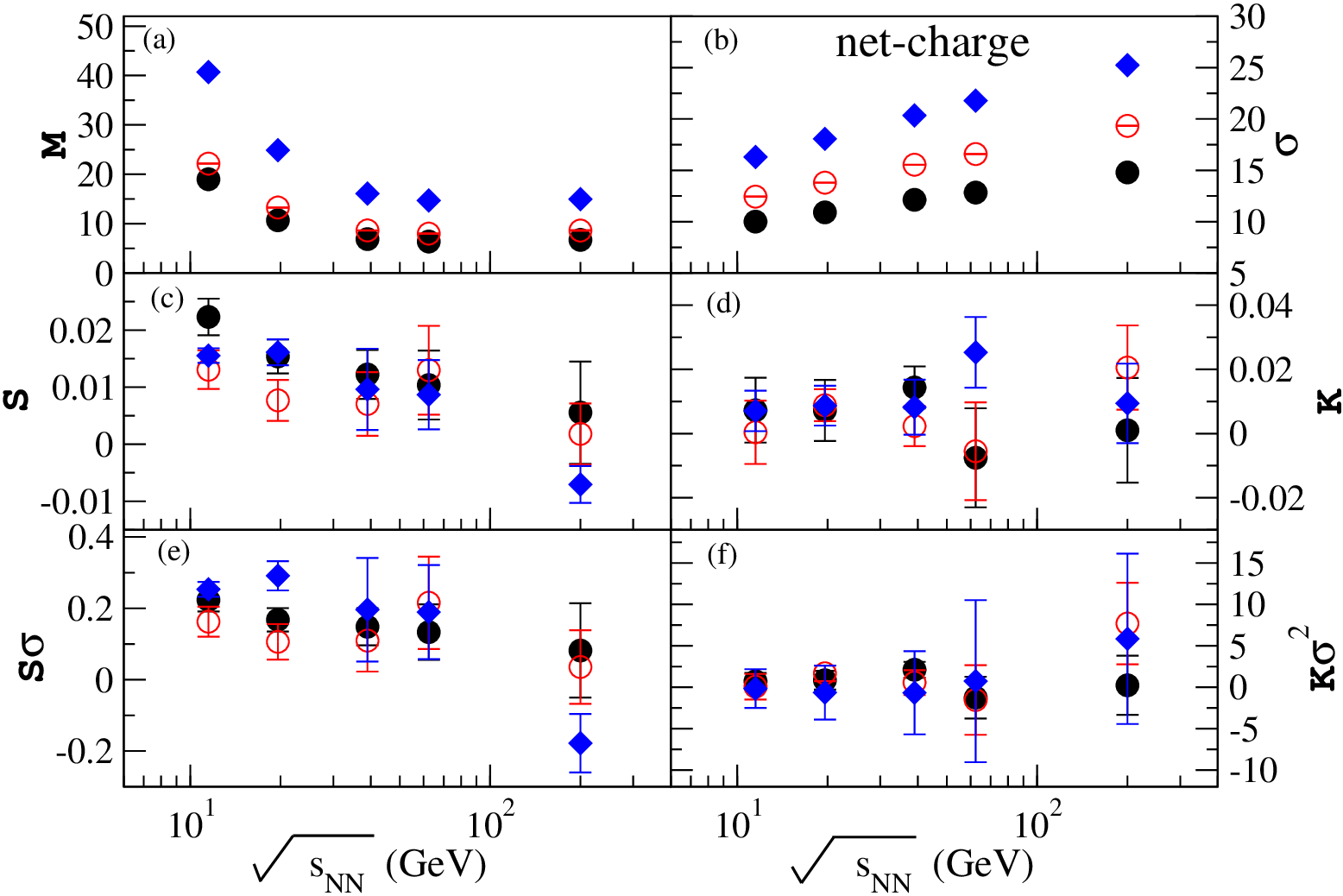}
\caption{(color online) Real higher moment excitation functions of
the net-proton (top), net-baryon (middle), and the net-charge
(bottom) in the 0-5\% most central Au+Au collisions calculated by
PACIAE for different window sizes.} \label{real_proton}
\end{figure}

\begin{figure}
\centering
\includegraphics[width=3.in]{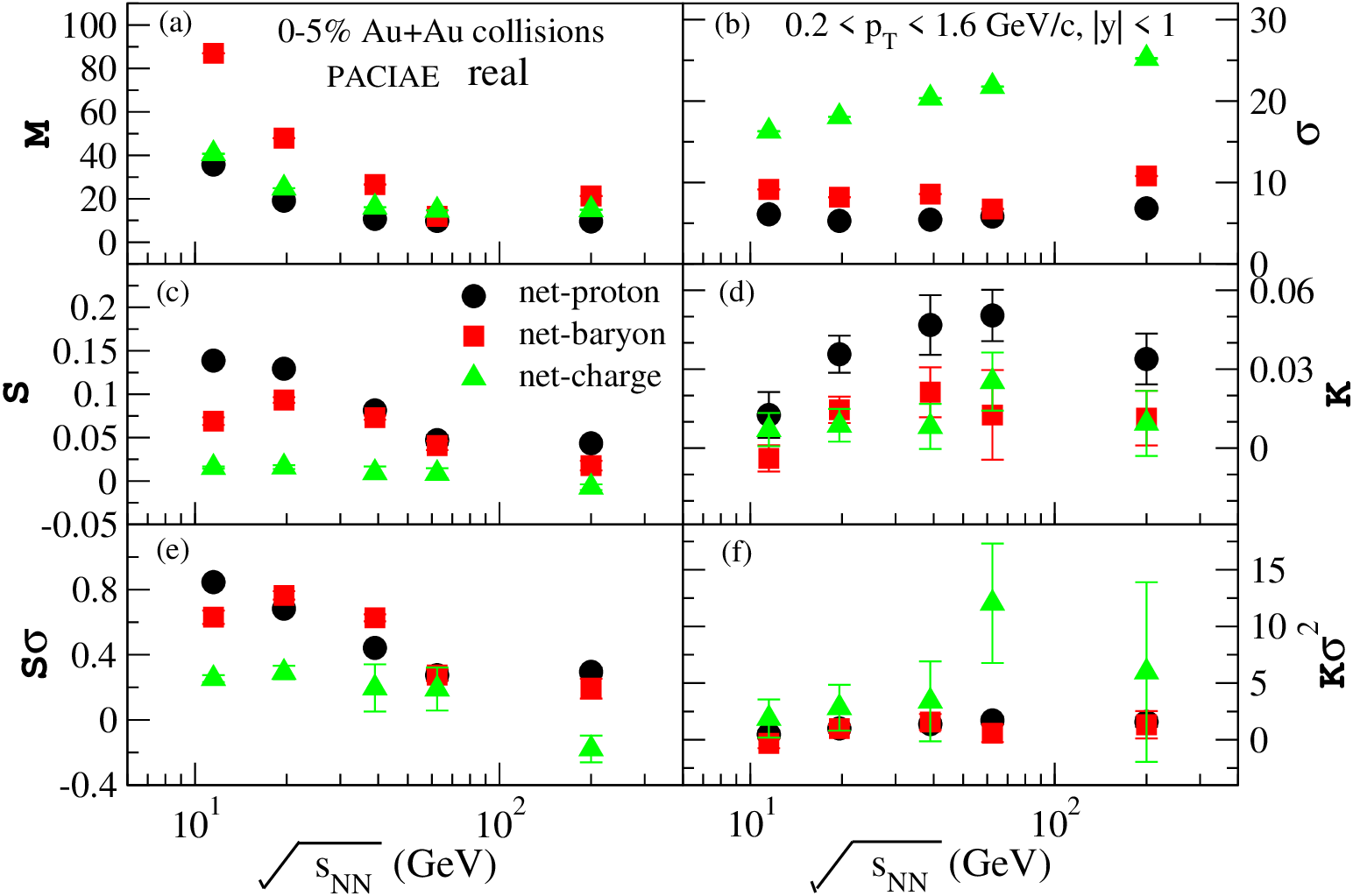}
\caption{(color online) Real higher moment excitation functions of
the net-proton, net-baryon, and the net-charge in the 0-5\% most
central Au+Au collisions calculated by PACIAE for a given window
size of $0.2 < p_T < 1.6$ GeV/c and $|y| < 1$.} \label{real_large}
\end{figure}

%%%%%%%%%%%%%%%%%%%%%%%%%%%%%%%%%%%%%%%
\section {Results}
%%%%%%%%%%%%%%%%%%%%%%%%%%%%%%%%%%%%%%%
We use PACIAE to generate the net-proton, net-baryon and the
net-charge number event distributions in 0-5\% most central Au+Au
collisions at $\sqrt {s_{NN}}$=11.5 to 200 GeV for the different
acceptance sizes. The real and non-statistical higher moment
excitation functions are then calculated, with all model parameters
fixed. Baryons are identified with $p + n + \Lambda$ and their
anti-particles (the non-strange baryons heavier than neutron and the
strange baryons heavier than $\Lambda$ are assumed decay already) and
the charge with $\pi^{\pm}$, $K^{\pm}$, $p$ and $\overline p$.

We summarize our results as follows:

Shown in the Figure \ref{real_proton}, from the top to the bottom, are
respectively the net-proton, net-baryon, and the net-charge real higher moment
excitation functions, with the mean $M$, the variance $\sigma$, skewness $S$,
kurtosis $\kappa$, and the products $S\sigma$ and $\kappa\sigma^2$ displayed
in the panels (a), (b), (c), (d), (e) and (f), respectively. In the figure,
the solid circles are the results in the acceptances of $0.4 < p_T < 0.8$
GeV/c and $|y|< 0.5$, the open circles are the results in $0.4 < p_T < 0.8$
and $|y| < 1$, and the diamonds are the results in $0.2 < p_T < 1.6$ and
$|y| < 1$.

Figure~\ref{real_large} gives, respectively, the net-proton (circles),
net-baryon (squares), and the net-charge (triangles) real higher moment
excitation functions for the mean (panel (a)), $\sigma$ ((b)), skewness ((c)),
kurtosis ((d)), $S\sigma$ ((e)), and $\kappa\sigma^2$ ((f)) calculated in the
$0.2 < p_T < 1.6$ GeV/c and $|y|<1$ acceptances.

We see from these figures that:
\begin{itemize}
\item
The shape of net-proton (net-baryon, net-charge) real higher moment
excitation function changing with the window sizes is approximately
similar only for the mean and $\sigma$, see top panel (a) and (b)
(middle panel (a) and (b), as well as bottom panel (a) and (b)) of
Fig.~\ref{real_proton}. This change becomes rather complicated for the
skewness $S$, kurtosis $\kappa$, and the product $S\sigma$ and $\kappa\sigma
^2$.
\item
We see in Fig.~\ref{real_proton} that the net-proton kurtosis is always
positive (cf. top panel (d)) unlike in \cite{blei}, the net-baryon kurtosis
decreases to a small negative value at low $\sqrt{s_{NN}}$ (cf. middle panel
(d)) but not to a lager negative value as mentioned in \cite{blei}, while the
net-charge kurtosis approximately keeps zero (cf. bottom panel (d)) like in
\cite{blei}.
\item
It is hard to see any signature, which may indicate the CP singularity,
from the real higher moment excitation functions shown in these
figures. Only the net-charge real excitation function of kurtosis
in the window of $0.4 < p_T < 0.8$ GeV/c and $|y| < 0.5$ as well as
$0.4 < p_T < 0.8$ GeV/c and $|y| < 1$ may indicate some signature (cf. bottom
panel (d) in Fig.~\ref{real_proton}).
\end{itemize}
\begin{figure}[h!]
%\vskip 7 mm
\centering
\includegraphics[width=3.in]{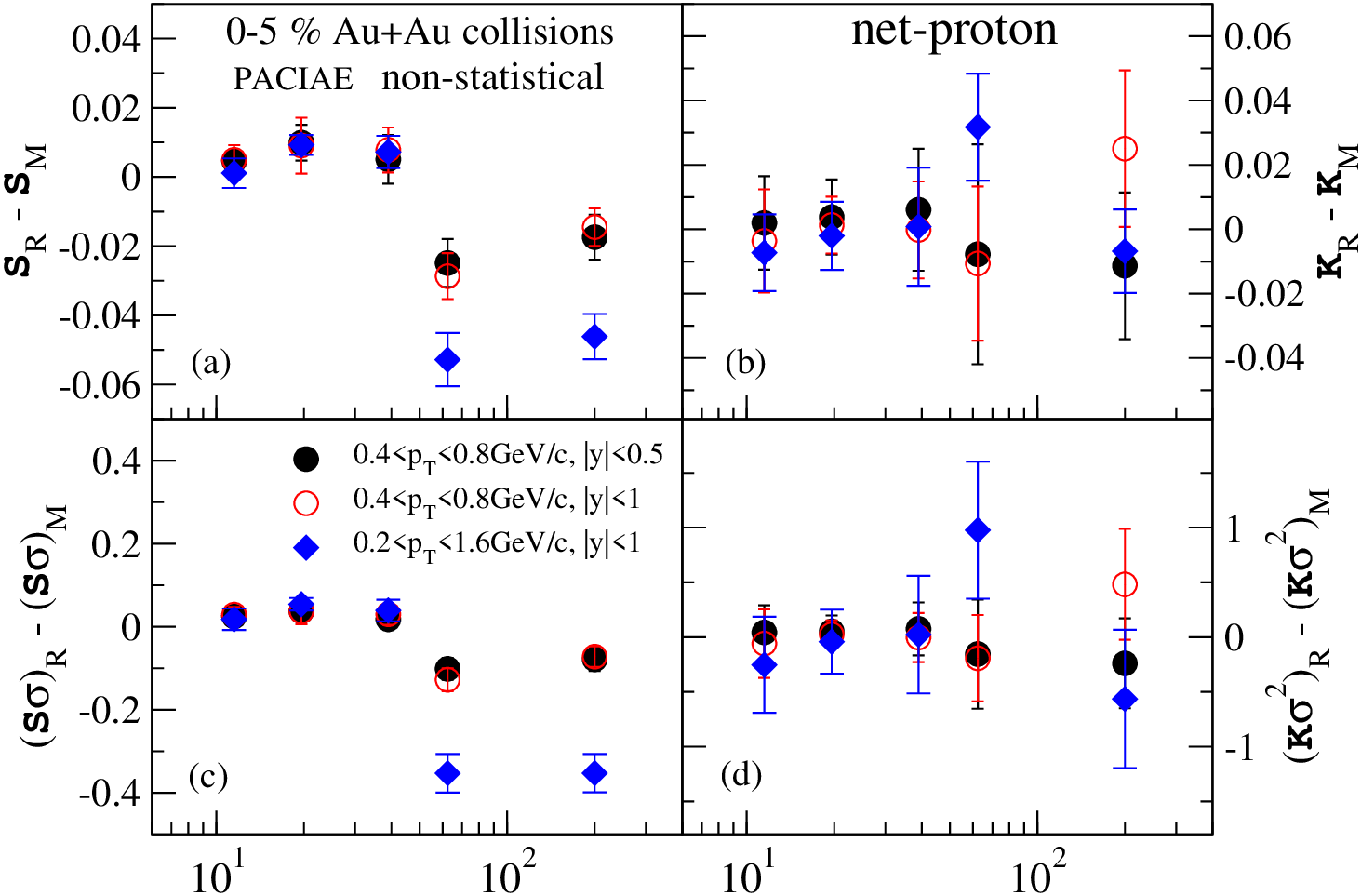}
\vskip 1.cm
\includegraphics[width=3.in]{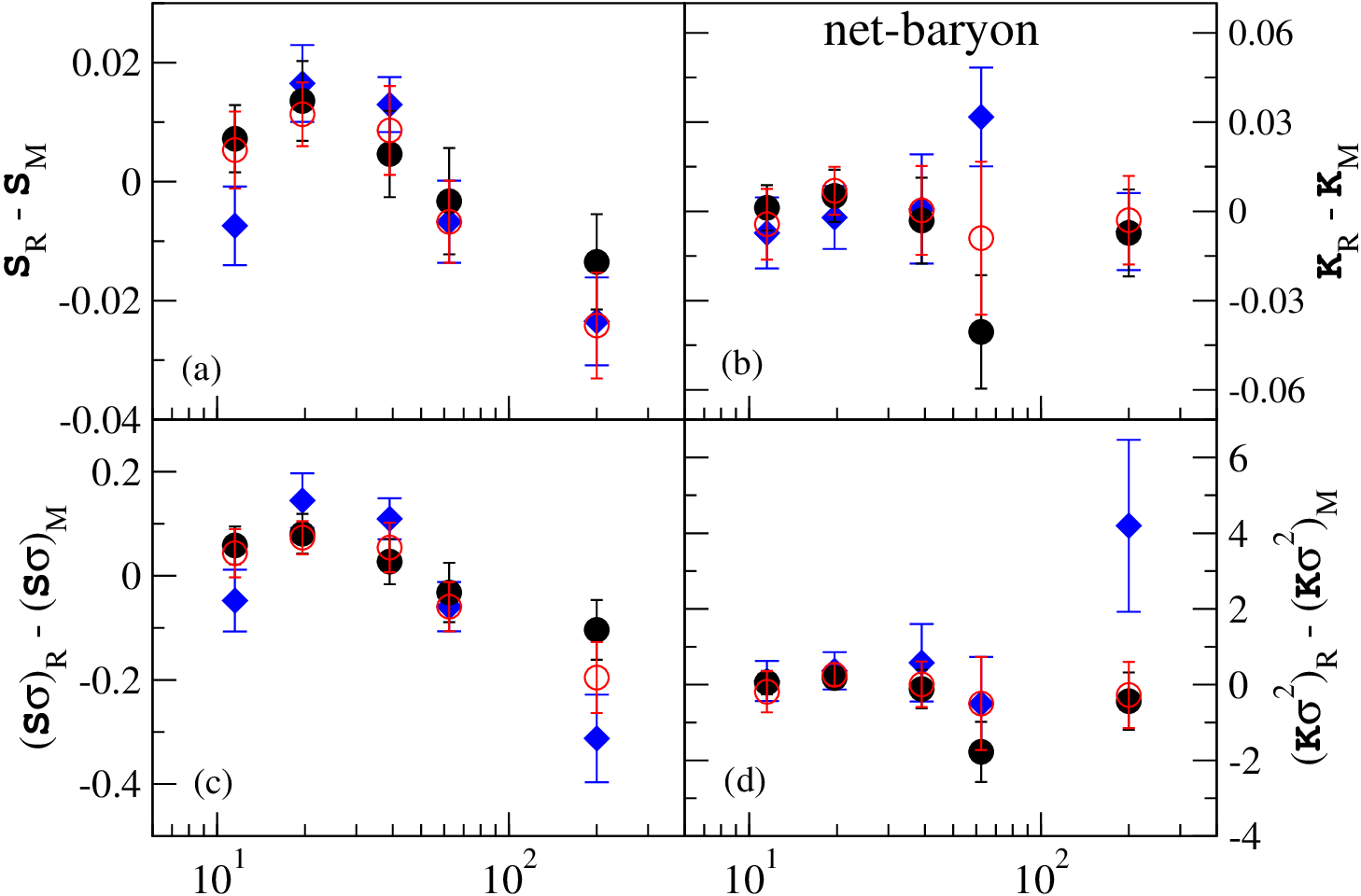}
\vskip 1.cm
\includegraphics[width=3.in]{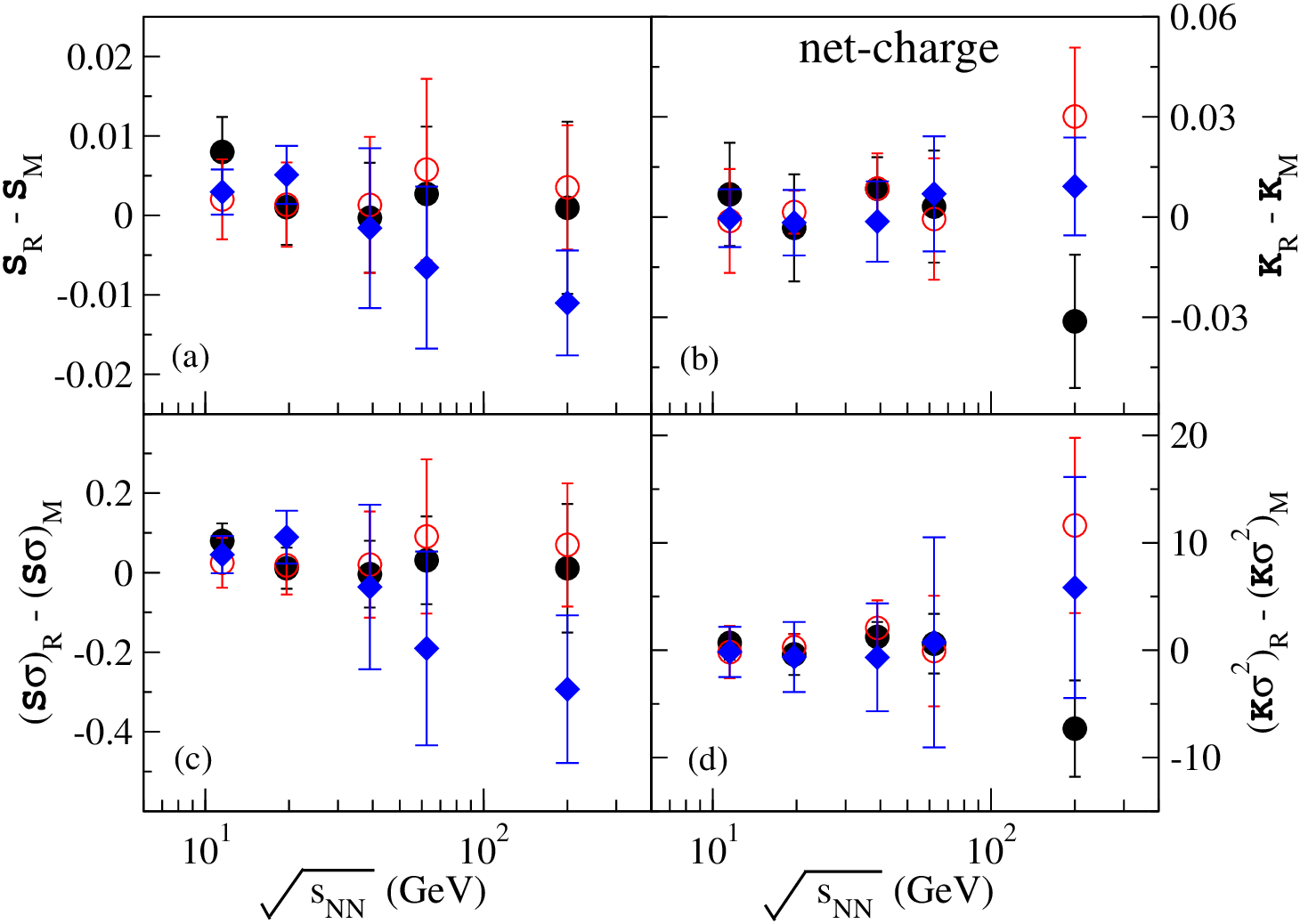}
\caption{(color online) Non-statistical higher moment excitation
functions of the net-proton (top), net-baryon (middle), and the
net-charge (bottom) in the 0-5\% most central Au+Au collisions
calculated by PACIAE for different window sizes.} \label{dfnetp}
\end{figure}

Figure~\ref{dfnetp} gives, from the top to the bottom, the
net-proton, net-baryon, and the net-charge non-statistical higher moment
excitation functions for the skewness (panel (a)), kurtosis (panel
(b)), $S\sigma$ (panel (c)), and $\kappa\sigma^2$ (panel (d)). In
the figure the solid circles, open circles, and the diamonds are,
respectively, the results in the acceptances of $0.4 < p_T < 0.8 $
GeV/c and $|y| < 0.5$, $0.4 < p_T < 0.8$ and $|y| < 1$, and $0.2 <
p_T < 1.6$ and $|y| < 1$. One sees an interesting structure, as
shown in the net-proton non-statistical excitation functions of the
skewness and $S\sigma$ (cf. top panels (a) and (c) in Fig.~\ref{dfnetp}),
which becomes more pronounced with increasing the window size. The structure
that the net-proton skewness and $S\sigma$ go down to a sizably negative
value within a small energy region may serve as a signature of the CP
singularity. As for the net-baryon non-statistical excitation functions, it
is the kurtosis and $\kappa\sigma^2$ which may display the CP singularity
only in the window of $0.4 < p_T <0.8$ GeV/c and $|y| < 0.5$, as shown in
the middle panels (b) and (d). However, one can not find any possible
signature of the CP singularity in the bottom panels.

\begin{figure}
\centering
\includegraphics[width=3.in]{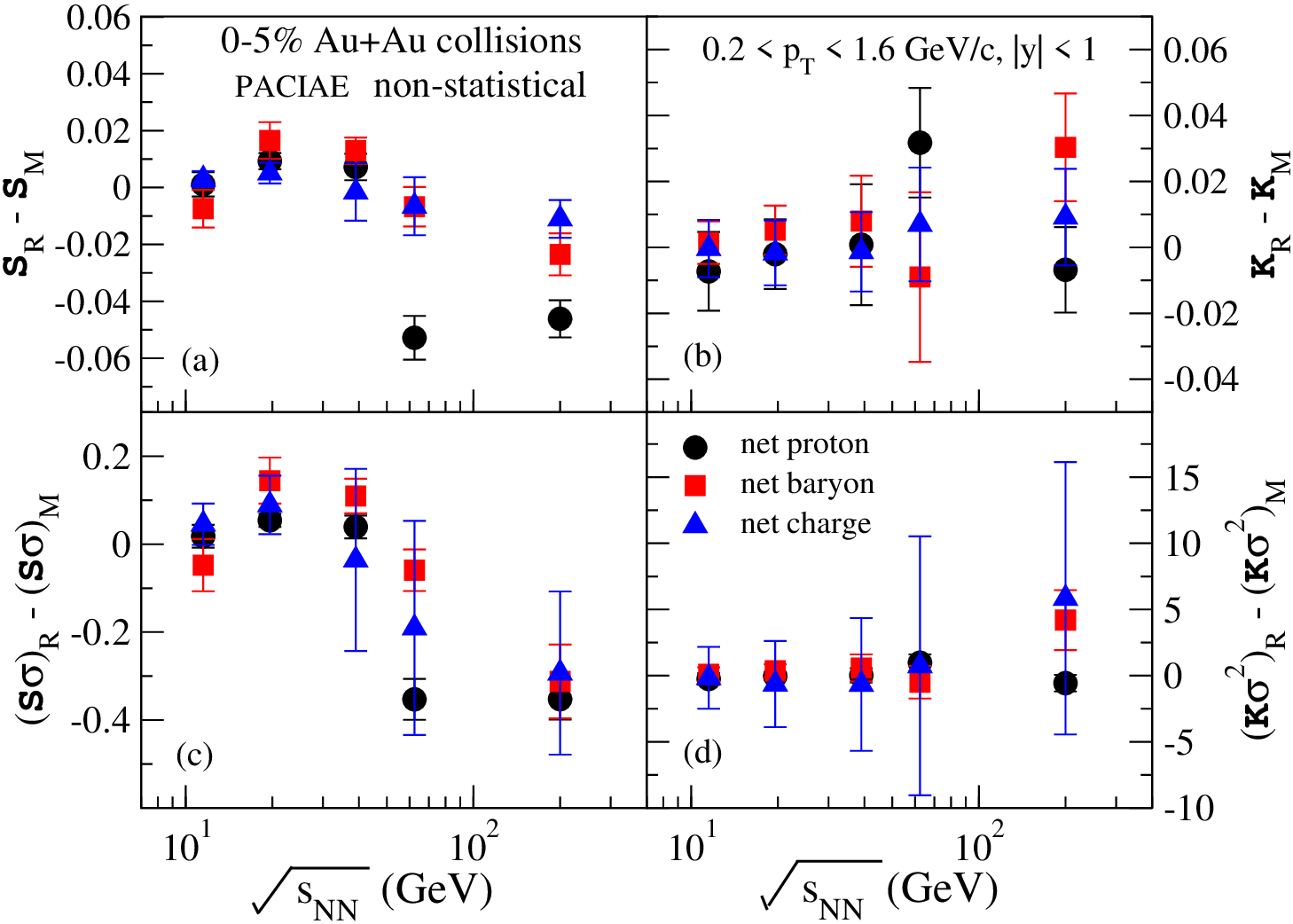}
\caption{(color online) Non-statistical higher moment excitation
functions of the net-proton (top), net-baryon (middle), and the
net-charge (bottom) in the 0-5\% most central Au+Au collisions
calculated by PACIAE for a given window sizes of $0.2 < p_T < 1.6$
GeV/c and $|y|<1$.} \label{dfnetpbch_large}
\end{figure}

Figure~\ref{dfnetpbch_large} gives, respectively, the net-proton
(circles), net-baryon (squares), and the net-charge (triangles)
non-statistical excitation functions for the skewness (panel (a)),
kurtosis (panel (b)), $S\sigma$ (panel (c)), and $\kappa \sigma^2$
(panel (d)) calculated in the acceptance of $0.2 < p_T < 1.6$ GeV/c
and $|y|<1$. We see in Fig.\ref{dfnetpbch_large} (a) and (c) that
the CP singularity may be shown in the non-statistical excitation
functions for the net-proton but not for the net-baryon and
net-charge. However, a CP signature may appear in the non-statistical
excitation functions of kurtosis for the net-baryon but not for the net-proton
and net-charge, as shown in Fig.\ref{dfnetpbch_large} (b).

%%%%%%%%%%%%%%%%%%%%%%%%%%%%%%%%%%%%%%%
\section {Conclusions}
%%%%%%%%%%%%%%%%%%%%%%%%%%%%%%%%%%%%%%%
We have calculated the real and non-statistical higher moment
excitation functions ($\sqrt{s_{NN}}$ = 11.5 to 200 GeV) for the
net-proton, net-baryon, and the net-charge number event
distributions in the relativistic Au+Au collisions with the parton
and hadron cascade model PACIAE \cite{yan}. The real event is
generated with the PACIAE model, while the mixed event is
constructed by combining particles selected from the real events
randomly. The non-statistical moment is derived as the difference
between real moment and the mixed moment. It turned out that because
of the statistical fluctuation dominance \cite{zhou} it is very hard
to see any signature of the CP singularity in the real higher moment
excitation functions of the net-proton, net-baryon, and the
net-charge, which is consistent with the results from the real
higher moment studies in \cite{star1,blei}.

The work suggests that the non-statistical moment of conserved
observables may play a potential role in exploring the CP
singularity in the relativistic heavy ion collisions. However, it is
found that the property of higher moment excitation functions is
significantly dependent on the window size. For a given conserved
observable, a signature of the CP singularity may appear in its
non-statistical higher moment excitation functions only in a
definite window size. But for a given widow size, the CP singularity
may be shown only in the non-statistical higher moment excitation
functions of a definite conserved observable as the multiplicity
distribution over rapidity and transverse momentum among different
conserved observables can be rather different.

Acknowledgements: This work was supported by the National Natural Science
Foundation of China under grant nos.: 10975062, 11075217, 11105227, 11175070,
11221504,11205066 and by the 111 project of the foreign expert bureau of China. 
AL and YPY acknowledge the financial support from TRF-CHE-SUT under contract No.
MRG5480186.

\end{document}